\documentclass[a4paper,journal,10pt]{IEEEtran}
\usepackage{amsmath,amsfonts}
\usepackage{algorithmic}
\usepackage{algorithm}
\usepackage{array}
\usepackage[caption=false,font=normalsize,labelfont=sf,textfont=sf]{subfig}
\usepackage{stfloats}
\usepackage{url}
\usepackage{verbatim}
\usepackage{graphicx}
\usepackage{cite}
\usepackage{color}
\usepackage{verbatim}
\usepackage{amssymb}
\usepackage{setspace}
\usepackage{booktabs}
\usepackage{graphicx}
\usepackage{float}
\usepackage{subfig}
\usepackage{bm}
\usepackage{multirow}
\usepackage{makecell}
\newtheorem{proposition}{Proposition}
\newcounter{tempEqCounter}
\begin{document}
	
\title{\LARGE Efficient Spectrum Sharing Between Coexisting OFDM Radar and Downlink Multiuser Communication Systems}	
\author{Jia Zhu, Yifeng Xiong, \IEEEmembership{Member, IEEE}, Junsheng Mu, Ronghui Zhang and Xiaojun Jing}
\markboth{}{}	

\maketitle	
\begin{abstract}
This paper investigates the problem of joint subcarrier and power allocation in the coexistence of radar and multi-user communication systems. Specifically, in our research scenario, the base station (BS) provides information transmission services for multiple users while ensuring that its interference to a separate radar system will not affect the radar's normal function. To this end, we propose a subcarrier and power allocation scheme based on orthogonal frequency division multiple access (OFDM). 
The original problem consisting involving multivariate fractional programming and binary variables is highly non-convex. 
Due to its complexity, we relax the binary constraint by introducing a penalty term, provided that the optimal solution is not affected. Then, by integrating multiple power variables into one matrix, the original problem is reformulated as a multi-ratio fractional programming (FP) problem, and finally a quadratic transform is employed to make the non-convex problem a sequence of convex problems. The numerical results indicate the performance trade-off between the multi-user communication system and the radar system, and notably that the performance of the communication system is not improved with power increase in the presence of radar interference beyond a certain threshold. 
This provides a useful insight for the energy-efficient design of the system.
\end{abstract}
	
\section{Introduction}
Radar-Communication Coexistence (RCC) has become the trend of future wireless system development, as both radar and communication systems develop towards higher frequency bands, larger antenna arrays and miniaturisation, becoming increasingly similar in hardware architectures, channel characteristics and signal processing \cite{hassanien2019dual,huang2015radar,cui2021integrating,liu2018mu}. Thus, both high quality wireless communication services and reliable radar sensing capabilities are ensured. This motivates the study of resource allocation (RA), especially the spectrum sharing between communication and radar systems \cite{9729765,9921194,10136774}.

Spectrum sharing requires a judicious allocation to mitigate interference and optimize resource utilization for RCC. Due to the difference in the functional purposes and performance metrics of communication and radar systems, there performance cannot be optimized using a unified utility function. Instead, one should maximize the performance of radar (resp. communication) system, under the constraints of ensuring communication (resp. radar) function and resource budget.

To accomplish the desired objectives, current research methods can be broadly classified into three categories. The first design strategy is a radar-centric design. It achieves coexistence between the two by limiting the interference of the radar system on the coexisting communication system \cite{aubry2015new,de2021joint}. In a similar vein, communication-centric approaches have been proposed in several recent studies, which aim for eliminating radar interference through a \textit{priori} knowledge or receiver design \cite{liu2018mimo,nartasilpa2018communications,wang2019power}. To this end, the third category involves jointly optimizing the coexisting systems to ensure that both communication and radar performance are satisfactory \cite{li2017joint,zheng2017joint,wang2019joint,yxtit2023}.

In this paper, we consider spectrum sharing between a single base station and a radar system using the OFDM system, where the base station serves multiple communication users simultaneously.
We have noticed that in related research on RCC spectrum sharing, either only spectrum resources are optimized or only power is optimized, and the existence of multi-users is often ignored in the case of joint optimization.
Unlike previous methods, our algorithm not only focuses on the performance trade-off between radar and communication systems, but also considers the multi-user situation. To this end, we design an effective and practical RA scheme to satisfy the radar system performance while maximizing the total multi-user sum rate. The main contributions of this algorithm are summarized as follows:
\begin{itemize}
	\item Different from the existing research work on spectrum sharing between radar and communication systems, we focus on resource allocation when multi-user communication systems and radar sensing coexist.
	\item The optimization problem we formulated is highly non-convex due to the inclusion of coupling variables and binary variables. We eliminate the binary variable using a continuous reformulation that yields identical solutions, and then transform the non-convex problem into a sequential convex problem by quadratic transformation.
	\item The numerical simulation results prove the effectiveness of the algorithm. Interestingly, we observe that under the radar performance constraint and interference, the communication sum rate does not improve as the allocated power grows beyond a certain threshold, this provides useful insights for energy-efficient design in practical applications.
\end{itemize}

The remainder of this paper is organized as follows. In Section \ref{Sec:System Descriptions}, we describe the system model and the optimization formulation. The subcarrier and power allocation algorithm is described in Section \ref{sec:Allocation Design for Sum Rate}. We evaluate the performance of the proposed algorithm by simulations in Section \ref{sec:Simulations Results}. Finally, we conclude the paper in Section \ref{sec:Conclusions}.

\section{System Descriptions and Problem Formulation}\label{Sec:System Descriptions}
\subsection{System Descriptions}
\begin{figure}[htbp]
	\centering
	\includegraphics[width=\columnwidth]{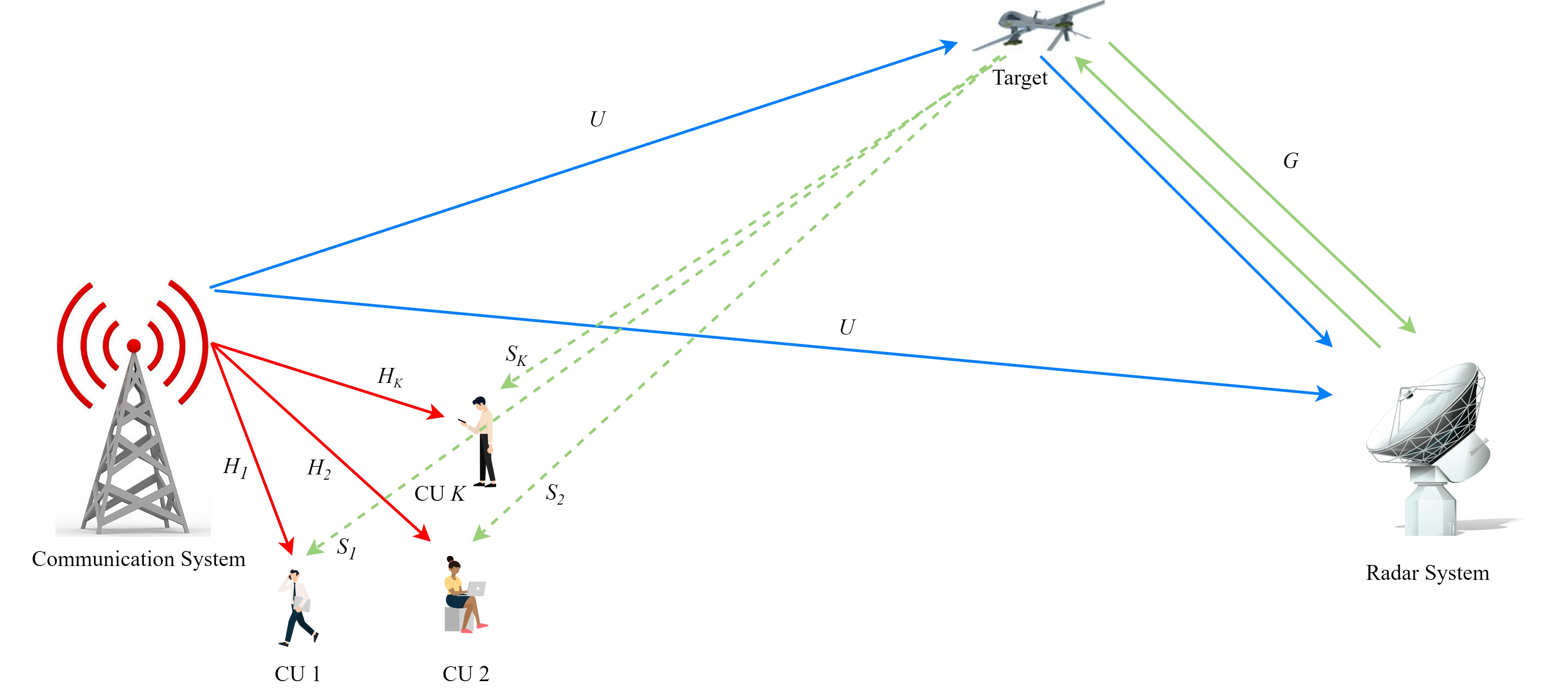}
	\caption{Diagram for co-existence of OFDM radar and downlink communication systems.}
	\label{fig-1}
\end{figure}
As depicted in \figurename~\ref{fig-1}, we consider a scenario where communication and radar coexist, in which both the communication system and the radar system employ OFDM waveforms with $N$ subcarriers. The BS provides service to $K$ downlink communication users (CUs).
The channels are assumed to be stationary over the observation period, and perfect channel state information for both the communication and radar channels is obtained in advance. The radar steers its beam at the potential target area according to the acquired \textit{a priori} knowledge, so the radar signal does not directly interfere with the CUs, but rather indirectly through target scattering.

For the downlink communication system, CUs not only receive communication signals from the BS, but also receive interference signals from radar system in the same frequency band. In particular, for CU $k$, the received signal can be represented as
\begin{equation}\label{CU-Signal}
	\bm{y}_{k}^c =\sum_{n=1}^{N}f_{n,k}\left( x_k P_n^c h_{n,k}^2 +x_rP_n^r s_{n,k}^2+m_{k}\right)
\end{equation}
where $f_{n,k}$ is the subcarrier sharing factor, $f_{n,k}=1$ indicates that subcarrier $n$ is assigned to CU $k$ and $f_{n,k} =0$ vice versa. $\bm{p}^c =[P_1^c, P_2^c, \dots,P_N^c]^T$ is the transmit power vector for communication system, $P_{n}^c$ is the power allocated to the subcarrier $n$. $\bm{p}^r =[P_1^r, P_2^r, \dots,P_N^r]^T$ is the transmit power vector for radar system, $P_{n}^r$ is the power allocated to the subcarrier $n$. $h_{n,k}$ is the channel gain from the BS to user $k$ on subcarrier $n$. $s_{n,k}$ is the interference channel gain from the radar transmitter to communication receiver $k$ on subcarrier $n$. $x_k$ is the symbols transmitted on subcarrier $n$ to CU $k$. The symbol streams $x_{k}$ are statistically independent with distribution $\mathcal{CN}(0,1)$. $x_r$ is the radar symbols transmitted on subcarrier $n$. The symbol streams $x^r$ are statistically independent with distribution $\mathcal{CN}(0,1)$. $m_{k}$ denotes the additive noise of the CU $k$. It is assumed to be distributed as $\mathcal{CN}(0,\sigma_{c,k}^2)$.

To this point, the achievable data rate of CU $k$ on subcarrier $n$ is given by 
\begin{equation}\label{subcarrier-rate-n}
	R_{n,k}=f_{n,k}\log_2\left(1+\frac{h_{n,k}^2P_{n}^c}{s_{n,k}^2P_{n}^r+\sigma_{c,k}^2}\right)
\end{equation}	
So we can get the total rate of CU $k$,
\begin{equation}\label{subcarrier-rate-k}
	R_{k}=\sum_{n=1}^{N}f_{n,k}\log_2\left(1+\frac{h_{n,k}^2P_{n}^c}{s_{n,k}^2P_{n}^r+\sigma_{c,k}^2}\right)
\end{equation}	
The data at the radar receiver with can be expressed as
\begin{equation}\label{Radar-Signal}
	\bm{y}_r=\sum_{n=1}^{N} (x_r P_n^r g_{n}^2+ \sum_{k=1}^{K}f_{n,k}x_k P_n^c u_{n}^2 +m_{r})
\end{equation}
where $g_n$ is the channel gain of radar system on subcarrier $n$, $u_n$ is the interference channel gain from the BS to radar receiver on subcarrier $n$. $m_r$ denotes the additive noise at the radar receiver. It is assumed to be distributed as $\mathcal{CN}(0,\sigma_{r}^2)$.

To ensure the normal operation of the radar function, we need to ensure that the signal-to-noise ratio (SINR) of the radar receiver is not lower than a certain specified threshold, 
\begin{equation}\label{SINR}
	\mathrm{SINR}= \frac{\sum_{n=1}^{N} g_n^2P_{n}^r}{\sum_{n=1}^{N}(\sum_{k=1}^{K}u_{n}^2f_{n,k}P_{n}^c+\sigma_r^2)} \ge \mu.
\end{equation}

To conclude, we have obtained the signal model for the communication system serving multiple users and the signal model for the radar system sensing a single target. Next, we formulated this problem as an optimization problem and then solved for its optimal solution.

\subsection{Optimization Problem Formulation}

We choose the sum rate of CUs as the optimization metric, while ensuring that the SINR of the radar system is above a preset threshold and satisfies the power constraint of the system, etc. The optimization problem is formulated as follows:
\begin{small}
    \begin{subequations}\label{Optimization Problem}
	\begin{align}
		&\max \ \  \sum_{k=1}^{K}\sum_{n=1}^{N}f_{n,k}\log_2\left(1+\frac{h_{n,k}^2P_{n}^c}{s_{n,k}^2P_{n}^r+\sigma_{c,k}^2}\right)\label{O-main} \\
		&\mathrm{s.t.}\ \ f_{n,k}\in \{ 0, 1\}, \forall n\in [1,2,\dots,N], \forall k\in [1,2,\dots,K]\label{O-1} \\
		&\quad \ \ \sum_{k=1}^{K} f_{n,k}\leq 1, \forall n \in [1,2,\dots,N] \label{O-2}\\
		&\quad \ \ \mathrm{SINR}\ge \mu ,  \label{SINR-Constraint}\\
		&\quad \ \ \sum_{k=1}^{K}\sum_{n=1}^{N} f_{n,k}P_{n}\leq P_{c}^{\max}, \label{O-4}\\
        &\quad \ \ \sum_{n=1}^{N} P_{n}^r\leq P_{r}^{\max}, \label{O-5}\\
        &\quad \ \ 0\leq P_{n}^c\leq P_c, \forall n\in  [1,2,\dots,N] \label{O-6}\\
        &\quad \ \ 0\leq P_{n}^r\leq P_r, \forall n\in  [1,2,\dots,N]\label{O-7}
	\end{align}
\end{subequations}
\end{small}
Constraints \eqref{O-1} and \eqref{O-2} ensure that each subcarrier is allocated to at most one CU, Constraint \eqref{SINR-Constraint} represents the minimum SINR for radar sensing. $P_{c}^{\max}$ in \eqref{O-4} and $P_{r}^{\max}$ in \eqref{O-5} are the maximum transmit powers of the communication and radar transmitters, respectively. Constraints \eqref{O-4} and \eqref{O-5} guarantee the transmit powers of communication and radar transmitters cannot go beyond their maximum limits. $P_c$ and $P_r$ represent the peak power constraints of communication subcarriers and radar subcarriers, respectively. It should be highlighted that constraints \eqref{O-6} and \eqref{O-7} has the effect of preventing the concentration of system power on one or a few subcarriers, thus avoiding the loss of frequency diversity advantage and the decrease of distance resolution in multi-carrier systems\cite{Sen_Tang_Nehorai_2010, Sit_Nuss_Zwick_2017}, as well as to prevent subcarrier interference caused by excessive peak power\cite{Papandreou_Antonakopoulos_2007}, which is practical and necessary.

Problem \eqref{Optimization Problem} is a mixed-integer non-convex optimization problem and seemingly intractable.
In particular, the non-convex combinatorial objective function \eqref{O-main}, the nonconvex constraint \eqref{SINR-Constraint} and the binary selection constraint \eqref{O-1} are the main obstacles for the design of the resource allocation algorithm.
Nevertheless, despite these challenges, in the next section, we will provide an efficient algorithm yielding near-optimal solution to problem \eqref{Optimization Problem}.

\section{Maximization Sum Rate based Allocation Design}\label{sec:Allocation Design for Sum Rate}
In this section, we reformulation the problem \eqref{Optimization Problem} by applying FP \cite{shen2018fractional}. Firstly, we relax the binary variable $f_{n,k}$ to a continuous variable and introduce a penalty term to ensure that the optimal solution of the \eqref{Optimization Problem} is not altered. Then, we merge the two type variables into one matrix and use FP to solve problem \eqref{Optimization Problem}.

\subsection{Equivalent continuous reformulation}
Firstly, an auxiliary variable $w_{n,k}=f_{n,k}P^c_{n} \in \left(0, P^c_{n}\right)$ is introduced to make the problem statement more concise. $w_{n,k} = P^c_{n}$ represents that subcarrier $n$ is allocated to CU $k$ and the corresponding power is $P^c_n$. By allowing $w_{n,k}$ to take continuous values in $\left(0, P^c_{n}\right)$, the communication rate \eqref{subcarrier-rate-n} may be rewritten as
\setcounter{equation}{6}
\begin{align}\label{subcarrier-rate-n-1}
		&R_{n,k}=\log_2\left(1+\frac{h_{n,k}^2 f_{n,k} P_n^c}{s_{n,k}^2P_{n}^r+\eta\sum_{i\neq k}^{K}h_{n,k}^2 f_{n,i} P_n^c+\sigma_{c,k}^2}\right)\nonumber \\
		&=\log_2\left(1+\frac{h_{n,k}^2 w_{n,k}}{s_{n,k}^2P_n^r+\eta\sum_{i\neq k}^{K}h_{n,k}^2 w_{n,i}+\sigma_{c,k}^2}\right),
\end{align}
where $\eta\sum_{i\neq k}^{K}h_{n,i}^2 w_{n,i}$ is a penalty term representing the interference term caused by subcarrier multiplexing. In particular, if the constraints \eqref{O-1} and \eqref{O-2} are satisfied, the value of the penalty term is zero. In fact, the optimal solutions of the relaxed problem always have zero penalty terms for appropriate choices of $\eta$, as indicated by the following proposition: 
\begin{proposition}\label{Proposition:equicvalence}
	Optimization problem \eqref{Optimization Problem} and \eqref{Optimization Problem re} are equivalent for all feasible solutions when $\eta \ge 1/2$.
 {\begin{small}
	\begin{subequations}\label{Optimization Problem re}
		\begin{align}
			&\max\limits_{w_{n,k}, P_n^r} \ \  \sum_{k=1}^{K}\sum_{n=1}^{N}\log_2\left(1+\frac{h_{n,k}^2 w_{n,k}}{s_{n,k}^2P_n^r+\eta\sum_{i\neq k}^{K}h_{n,k}^2 w_{n,i}+\sigma_{c,k}^2}\right) \label{O-main-re} \\
			&\mathrm{s.t.}\ \ \mathrm{SINR} \ge \mu \label{O-re-1}\\
			&\quad \ \ \sum_{k=1}^{K}\sum_{n=1}^{N} w_{n,k}\leq P_{c}^{\max}, \label{O-re-2}\\
			&\quad \ \ \sum_{n=1}^{N} P_n^r\leq P_{r}^{\max} \label{O-re-3}\\
			&\quad \ \ 0\leq w_{n,k}\leq P_c, \forall n\in  [1,2,\dots,N], \forall k\in  [1,2,\dots,K] \label{O-re-4}\\
			&\quad \ \ 0\leq P_n^r\leq P_r, \forall n\in  [1,2,\dots,N] \label{O-re-5}
		\end{align}
	\end{subequations}
  \end{small}}
\end{proposition}
\begin{IEEEproof}
	Assume the total communication transmit power allocated to subcarrier $n$ is $W_{n}$ and $\sum_{k=1}^K w_{n,k}=W_{n}$. Denote $\delta_{n,k} \triangleq \sum_{i\neq k}^K w_{n,i}$ represent the power allocated to other subcarriers. The communication rate of user $k$ on subcarrier $n$ in \eqref{subcarrier-rate-n} can be rewritten as
	\begin{align}
		R_{n,k} & =\log _2\left(1+\frac{h_{n,k}^2 w_{n,k}}{s_{n,k}^2 P^r_{n}+\eta h_{n,k}^2 \delta_{n,k}+\sigma_{c,k}^2}\right) \nonumber \\
		& =\log _2\left(1+\frac{W_n-\delta_{n,k}}{s_{n,k}^2h_{n,k}^{-2}P^r_{n}+\eta \delta_{n,k}+\sigma_{c,k}^2 h_{n,k}^{-2}}\right) .
	\end{align}
 
Let us first consider the scenario that there are only two users ($K=2$). We are interested in the condition under which the following holds
\begin{align}\label{11}
    \log_2 \left(1\!+\!\frac{W_n}{\zeta_{n,1}}\right)\!\ge&\! \log_2 \left(1+\frac{W_n - \delta_{n,1}}{\zeta_{n,1}\!+\!\eta \delta_{n,1}}\right) 
        \!+ \nonumber \\
        &\!\log_2 \left(1\!+\!\frac{\delta_{n,1}}{\zeta_{n,2}+\eta(W_n -\delta_{n,1})} \right),
\end{align}
namely that it is better not to share the power between the two users, where $\zeta_{n,1}=\frac{s_{n,1}^2}{h_{n,1}^2}P^r_{n}+\frac{\sigma_{c,1}^2}{h_{n,1}^2}$ and $\zeta_{n,2}=\frac{s_{n,2}^2}{h_{n,2}^2}P^r_{n}+\frac{\sigma_{c,2}^2}{h_{n,2}^2}$, and $\zeta_{n,1}\leq \zeta_{n,2}$.

After some algebra, we see that \eqref{11} holds whenever \eqref{10} holds. 

The equality is apparently achieved when $\delta_{n,1}=0$. Next, we wish to investigate the condition under which \eqref{10} holds for all $\delta_{n,1}\geq 0$. To this end, it suffices to show that $f^{\prime}(\delta_{n,1})\leq 0,~\forall \delta_{n,1}\geq 0$. Taking the derivative of $f(\delta_{n,1})$ with respect to $\delta_{n,1}$, we have \eqref{18}.

Through observation, we can determine that \eqref{leq} holds.

In other words, as long as $$\frac{-2\eta (W_n-\delta_{n,1})+2\eta \delta_{n,1}+W_n-2\delta_{n,1} -\zeta_{n,2}+\zeta_{n,1}}{(\eta\delta_{n,1}+\zeta_{n,1})(\zeta_{n,1}(W_n-\delta_{n,1})+\zeta_{n,2})}\leq 0,$$ the condition $f^{\prime}(\delta_{n,1})\leq 0, ~\forall \delta_{n,1}\geq 0$ will certainly be satisfied.

Thus, we see that a sufficient condition for $f^{\prime}(\delta_{n,1})\leq 0, ~\forall \delta_{n,1}\geq 0$ is
\setcounter{tempEqCounter}{\value{equation}}
\setcounter{equation}{13}
\begin{equation}\label{eta}
    \eta \ge \frac{1}{2}+\frac{\zeta_{n,1}-\zeta_{n,2}}{W_n-2\delta_{n,1}}.
\end{equation}
Since the term $\frac{\zeta_{n,1}-\zeta_{n,2}}{W_n-2\delta_{n,1}}\leq 0$, we may conclude that $\eta \ge \frac{1}{2}$ is sufficient for \eqref{11} to hold for any $\delta_{n,1}\geq 0$. 

We may extend the result to the case of $K=3$. 
By viewing $W_n-\delta_{n,1}$ as the total power (denoted by $\widetilde{W}_n$), we are ainterested in the condition under which the following holds
\setcounter{tempEqCounter}{\value{equation}}
\setcounter{equation}{14}
\begin{align}\label{20}
        \log_2 \left(1\!+\!\frac{\widetilde{W}_n}{\widetilde{\zeta}_{n,1}}\right)\!\ge &\! \log_2 \left(1+\frac{\widetilde{W}_n-\delta_{n,2}}{\widetilde{\zeta}_{n,1}\!+\!\eta \delta_{n,2}}\right) 
        \!+\\ \nonumber
        &\!\log_2 \left(1\!+\!\frac{\delta_{n,2}}{\widetilde{\zeta}_{n,2}+\eta(\widetilde{W}_n -\delta_{n,2})} \right),
\end{align}
where $\zeta_{n,1}+(1/2)\delta_{n,1}$ and $\zeta_{n,2}+(1/2)\delta_{n,1}$ as the noise plus interference (denoted by $\widetilde{\zeta}_{n,1}$ and $\widetilde{\zeta}_{n,2}$, respectively. $\widetilde{\zeta}_{n,1}\leq \widetilde{\zeta}_{n,2}$.)

After some algebra, we see that \eqref{20} holds whenever \eqref{21} holds. 

Similar to the case when $K=2$, the equality is apparently achieved when $\delta_{n,2}=0$. Next, we wish to investigate the condition under which \eqref{21} holds for all $\delta_{n,2}\geq 0$. To this end, it suffices to show that $f^{\prime}(\delta_{n,2})\leq 0, ~\forall \delta_{n,2}\geq 0$. Taking the derivative of $f(\delta_{n,2})$ with respect to $\delta_{n,2}$, we can derive an inequality equivalent
\setcounter{equation}{16}
\begin{equation}\label{eta2}
    \eta \ge \frac{1}{2}+\frac{\widetilde{\zeta}_{n,1}-\widetilde{\zeta}_{n,2}}{\widetilde{W}_n-2\delta_{n,2}}.
\end{equation}
Since the term $\frac{\widetilde{\zeta}_{n,1}-\widetilde{\zeta}_{n,2}}{\widetilde{W}_n-2\delta_{n,2}}\leq 0$, we may conclude that $\eta \ge \frac{1}{2}$ is sufficient for \eqref{20} to hold for any $\delta_{n,2}\geq 0$. 

By employing the method of mathematical induction, the previous arguments can be reused to show that allocating power to $(k+1)$ users is never better than all strategies that allocate power to $k$ users, and hence allocating power exclusively to a single user is always the optimal choice. 

As a conclusion, the optimization problem \eqref{Optimization Problem} and \eqref{Optimization Problem re} are equivalent whenever $\eta \geq 1/2$.
\end{IEEEproof}

\setcounter{tempEqCounter}{\value{equation}}
\setcounter{equation}{10}
\begin{figure*}[tbp]
\begin{equation}\label{10}
    \frac{W_n}{\zeta_{n,1}}\ge f(\delta_{n,1})=\frac{\delta_{n,1}(W_n-\delta_{n,1})+(W_n-\delta_{n,1})(\zeta_{n,2}+\eta(W_n-\delta_{n,1}))+\delta_{n,1}(\zeta_{n,1}+\eta\delta_{n,1})}{(\zeta_{n,1}+\eta\delta_{n,1})(\zeta_{n,2}+\eta(W_n-\delta_{n,1}))}
\end{equation}
\end{figure*}
\setcounter{equation}{11}
\begin{figure*}[tbp]
\begin{equation}\label{18}
\begin{split}
	f^{\prime}(\delta_{n,1})= & \frac{-2\eta (W_n-\delta_{n,1})+2\eta \delta_{n,1}+W_n-2\delta_{n,1} -\zeta_{n,2}+\zeta_{n,1}}{(\eta\delta_{n,1}+\zeta_{n,1})(\zeta_{n,1}(W_n-\delta_{n,1})+\zeta_{n,2})} +\\
	 &\frac{\eta\left(\eta(W_n-\delta_{n,1})^2+\delta_{n,1}(\eta \delta_{n,1}+\zeta_{n,1})+\zeta_{n,2}(W_n-\delta_{n,1})+\delta_{n,1}(W_n-\delta_{n,1})\right)}{(\eta \delta_{n,1}+\zeta_{n,1})(\eta(W_n-\delta_{n,1})+\zeta_{n,2})^2}-\\
	 &\frac{\eta\left(\eta(W_n-\delta_{n,1})^2+\delta_{n,1}(\eta \delta_{n,1}+\zeta_{n,1})+\zeta_{n,2}(W_n-\delta_{n,1})+\delta_{n,1}(W_n-\delta_{n,1})\right)}{(\eta \delta_{n,1}+\zeta_{n,1})^2(\eta(W_n-\delta_{n,1})+\zeta_{n,2})}
\end{split}
\end{equation}
\end{figure*} 

\begin{figure*}[tbp]
\begin{equation}\label{leq}
\begin{split}
	 &\frac{\eta\left(\eta(W_n-\delta_{n,1})^2+\delta_{n,1}(\eta \delta_{n,1}+\zeta_{n,1})+\zeta_{n,2}(W_n-\delta_{n,1})+\delta_{n,1}(W_n-\delta_{n,1})\right)}{(\eta \delta_{n,1}+\zeta_{n,1})(\eta(W_n-\delta_{n,1})+\zeta_{n,2})^2}-\\
	 &\frac{\eta\left(\eta(W_n-\delta_{n,1})^2+\delta_{n,1}(\eta \delta_{n,1}+\zeta_{n,1})+\zeta_{n,2}(W_n-\delta_{n,1})+\delta_{n,1}(W_n-\delta_{n,1})\right)}{(\eta \delta_{n,1}+\zeta_{n,1})^2(\eta(W_n-\delta_{n,1})+\zeta_{n,2})} \leq 0.
\end{split}
\end{equation}
\end{figure*}
\setcounter{equation}{15}
\begin{figure*}[tbp]
\begin{equation}\label{21}
    \frac{\widetilde{W}_n}{\widetilde{\zeta}_{n,1}}\ge f(\delta_{n,2})=\frac{\delta_{n,2}(\widetilde{W}_n-\delta_{n,2})+(\widetilde{W}_n-\delta_{n,2})(\widetilde{\zeta}_{n,2}+\eta(\widetilde{W}_n -\delta_{n,2}))+\delta_{n,2}(\widetilde{\zeta}_{n,1}+\eta\delta_{n,2})}{(\widetilde{\zeta}_{n,1}+\eta\delta_{n,2})(\widetilde{\zeta}_{n,2}+\eta(\widetilde{W}_n-\delta_{n,2}))}
\end{equation}
{\noindent} \rule[-10pt]{18cm}{0.05em}
\end{figure*}

\subsection{Sequential convex relaxation}
Although we have relaxed the binary variable into continuous variable, the existence of coupling variables $w_{n,k}$ and $P^r_n$ in \eqref{Optimization Problem re} makes it still a non-convex problem. 
To solve \eqref{Optimization Problem re}, alternating optimization is a common solution method. By fixing one variable and optimizing another variable, the original problem is decomposed into two sub-problems. The disadvantage of this method is that the decomposed sub-problem is still a non-convex optimization problem, which has high computational complexity and is difficult to obtain the optimal solution. Inspired by \cite{WangPowerAllocation}, we combine the variables $w_{n,k}$ and $P^r_n$ to be optimized into matrix variable $\bm{P}$, avoiding the process of alternating optimization, and only need to update the matrix variables to get the solution of the problem.

Specifically, we define $\bm{e}_k =[\bm{0}_{k-1}; 1; \bm{0}_{K+1-k}]^T$, $\bm{\alpha}_{n,k}=\frac{h_{n,k}^2}{\sigma_{c,k}^2}\bm{e}_k$, $\bm{\beta}_{n,k}=\frac{s_{n,k}^2}{\sigma_{c,k}^2}\bm{e}_k$, $\bm{\xi}_n = \frac{g_n^2}{\sigma_r^2}\bm{e}_{K+1}$, $\bm{\gamma}_{n}=[\frac{u_n^2}{\sigma_r^2},\dots,\frac{u_n^2}{\sigma_r^2}, 0]^T$ and $\bm{P} = [\bm{w}_1;\bm{w}_2;\dots;\bm{w}_K;\bm{p}^r]$ is a $(K+1)\times N$ matrix, $\bm{w}_k =[w_{1,k},w_{2,k},\dots,w_{N,k}]^T$. we rewrite \eqref{subcarrier-rate-n-1} and \eqref{SINR} as
\setcounter{equation}{17}
\begin{small}
    \begin{equation}\label{R_k}	R_k(\bm{P})\!=\!\!\sum_{n=1}^{N}\log_2\!\left(\!1\!+\!\frac{\bm{\alpha}_{n,k}^T\bm{P}\bm{v}_n}{\bm{\beta}_{n,k}^T\bm{P}\bm{v}_n\!+\!\eta\!\sum_{i\neq k}^{K}\!\bm{\alpha}_{n,i}^T\bm{P}\bm{v}_n\!+\!1}\!\right),
\end{equation}
\end{small}
\begin{equation}\label{SINR-Re}
	\mathrm{SINR}(\bm{P})= \frac{\sum_{n=1}^{N} \bm{\xi}_n^T\bm{P}\bm{v}_n}{\sum_{n=1}^{N}(\bm{\gamma}_n^T\bm{P}\bm{v}_n+1)},
\end{equation}
where $\bm{v}_n$ is $N$-dimensional vector, $\bm{v}_n(j) = 1$ when $j=n$ and $\bm{v}_n(j) = 0$ otherwise.
Then, \eqref{Optimization Problem re} can be rewritten as
\begin{subequations}\label{OP1}
	\begin{align}
		&\max\limits_{\bm{P}} \ \  \sum_{k=1}^{K} R_k(\bm{P}) \\
		&\mathrm{s.t.}\quad \eqref{SINR-Re}, \eqref{O-re-2}-\eqref{O-re-5} 
	\end{align}
\end{subequations}

Problem \eqref{OP1} remains a challenging non-convex problem due to the strong interdependence of the transmit power levels of different subcarriers, as reflected in the interference terms of the SINR. We take the quadratic transform is proposed in \cite{shen2018fractional} to address the multiple-ratio FP problems.
By performing a quadratic transform on each SINR term, we obtain the following reformulation
\begin{subequations}\label{OP2}
	\begin{align}
		&\max\limits_{\bm{P},\bm{Y}} \ \  Q(\bm{P},\bm{Y}) \\
		&\mathrm{s.t.}\quad \eqref{SINR-Re}. \eqref{O-re-2}-\eqref{O-re-5} 
	\end{align}
\end{subequations}
where
\begin{equation}
	\begin{aligned}
		& Q(\bm{P},\bm{Y})=\sum_{k=1}^{K} \sum_{n=1}^{N}\log_2 \left(1+2 y_{n,k} \sqrt{\bm{\alpha}_{n,k} \bm{P}\bm{v}_n}\right. \\
		&\left.-y_{n,k}^2\left(\bm{\beta}_{n,k}\bm{P}\bm{v}_n+\eta\sum_{i \neq k}^{K}\bm{\alpha}_{n,i} \bm{P}\bm{v}_n+1\right)\right),
	\end{aligned}
\end{equation}
where $y_{n,k} =[\bm{Y}]_{n,k}$ is the auxiliary variable introduced by the quadratic transform for each CU $k$ on subcarrier $n$.

We update $y_{n,k}$ and $\bm{P}$ in an iterative fashion. The optimal $y_k$ for fixed $\bm{P}$ is
\begin{equation}\label{y}
	y_{n,k}^*=\frac{\sqrt{\bm{\alpha}_{n,k}^T\bm{P}\bm{v}_n}}{\bm{\beta}_{n,k}^T\bm{P}\bm{v}_n+\eta\sum_{i\neq k}^{K}\bm{\alpha}_{n,i}^T\bm{P}\bm{v}_n+1}
\end{equation}
Then, finding the optimal $\bm{P}$ for fixed $y_{n,k}$ is a convex problem and can be solved by off-the-shelf convex optimization solvers. 
\begin{flushleft}
	\begin{spacing}{0.75}
		\begin{tabular*}{\hsize}{@{}@{\extracolsep{\fill}}l@{}} 
			\toprule
			\specialrule{0em}{2pt}{2pt}
			\textbf{Algorithm 1:} $\textbf{Joint Design Algorithm}$   \\
			\midrule
			\specialrule{0em}{2pt}{2pt}
			\specialrule{0em}{1pt}{1pt}
			\textbf{Input:} $h_{n,k}$, $s_{n,k}$, $u_n$, $g_n$, $\eta$, $m_k$, $m_r$, $\bm{p}^c$, $\bm{p}^r$.\\
			\specialrule{0em}{1pt}{1pt}
		    \textbf{Output:} Communication power $\bm{p}^c$, Radar power $\bm{p}^r$.\\
			\specialrule{0em}{1pt}{1pt}
			\textbf{Initialization:} Initialize $\bm{p}^c$, $\bm{p}^r$ and $\eta$ to feasible values.\\
			\specialrule{0em}{1pt}{1pt}
			\textbf{$\textbf{Repeat}$} \\
			\specialrule{0em}{1pt}{1pt}
			\qquad \textbf{$\textbf{1}$.} Solve Problem \eqref{y}. \\ 
			\specialrule{0em}{1pt}{1pt}
			\qquad \textbf{$\textbf{2}$.} Update $\bm{P}$ by solving the reformulated\\ 
			\quad\qquad  convex optimization problem \eqref{OP2} for fixed $y_{n,k}$.\\
			\specialrule{0em}{1pt}{1pt}
			\textbf{until} convergence.\\
			\bottomrule
		\end{tabular*} 
	\end{spacing}
\end{flushleft}


\section{Simulations Results}\label{sec:Simulations Results}
We consider a scenario where one BS serves 5 CUs are randomly distributed within the cell. The main simulation parameters are listed in table I.

\begin{table}[t] \label{table1} 
	\setlength{\abovecaptionskip}{0.05cm} 
	\centering
	\caption{Simulation Parameters} 
        \scalebox{0.8}{
	\begin{tabular*}{\hsize}{@{\extracolsep{\fill}}l l c c} 
		\toprule
		Parameters & Values \\
		\midrule
		Number of subcarriers & $128$  \\
		Carrier frequency & $2.4$ GHz  \\
		Cell radius & $800$ m  \\
		noise variance $\sigma_{c,k}^2$& $-105$ dB\\
		noise variance $\sigma_{r}^2$& $-105$ dB\\
		Maximum transmit power $P_c^{\max}$& $50$ dBm\\
		Maximum transmit power $P_r^{\max}$& $45$ dBm\\
		Maximum subcarrier power $P_c$& $30$ dBm\\
		Maximum subcarrier power $P_r$& $30$ dBm\\
		Shadowing distribution & Log-normal\\
		Shadowing standard deviation & $8$ dB\\
		Pathloss model & WINNER II \cite{bultitude20074}\\
		\bottomrule
	\end{tabular*}}
\end{table}

\begin{figure}[t]
	\centering
	\includegraphics[width=0.85\columnwidth]{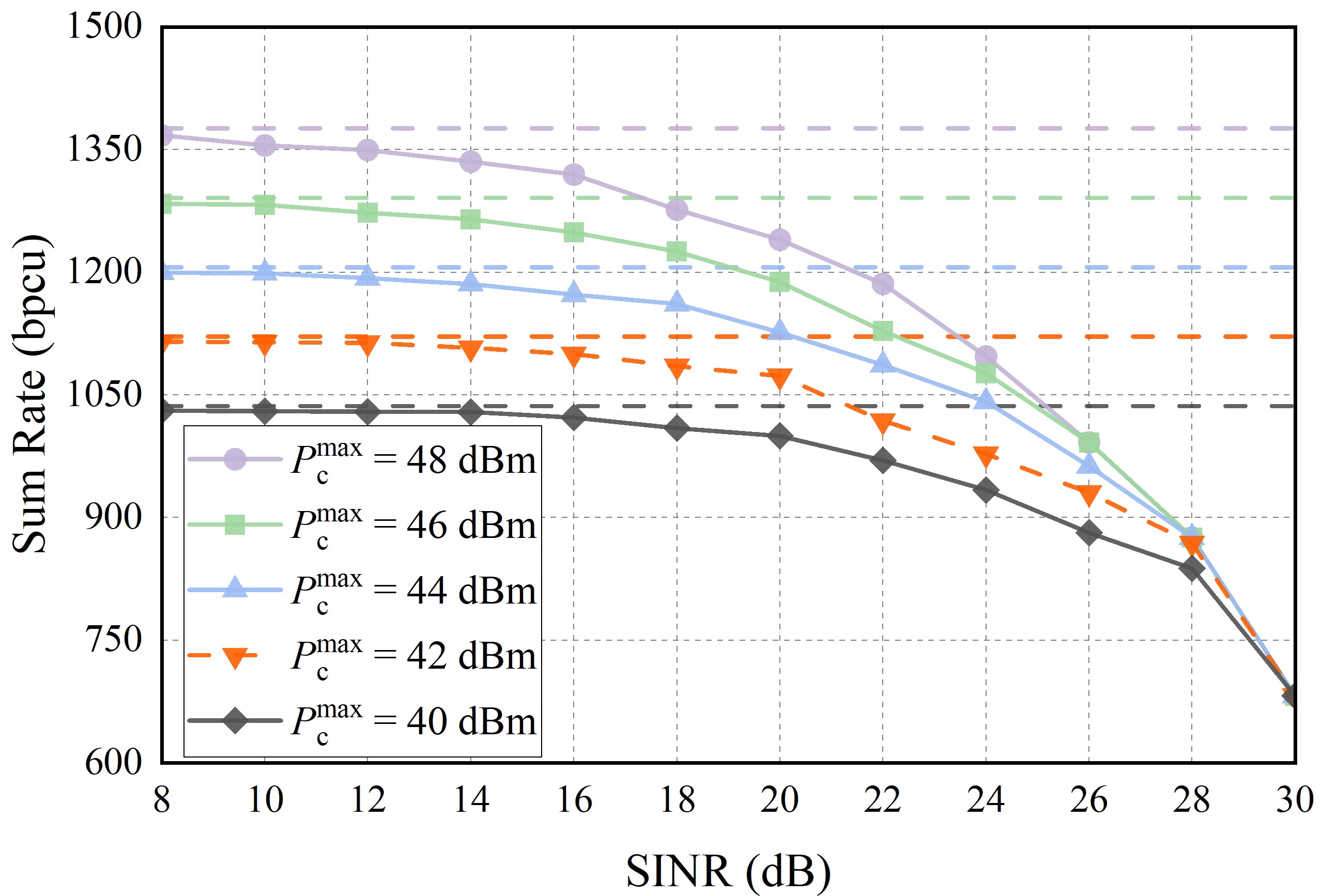}
	\caption{Sum rate versus SINR under different values of $P_c^{\max}$}
    \vspace{-3mm}
	\label{fig-2}
\end{figure}

\begin{figure}[t]
	\centering
	\includegraphics[width=0.85\columnwidth]{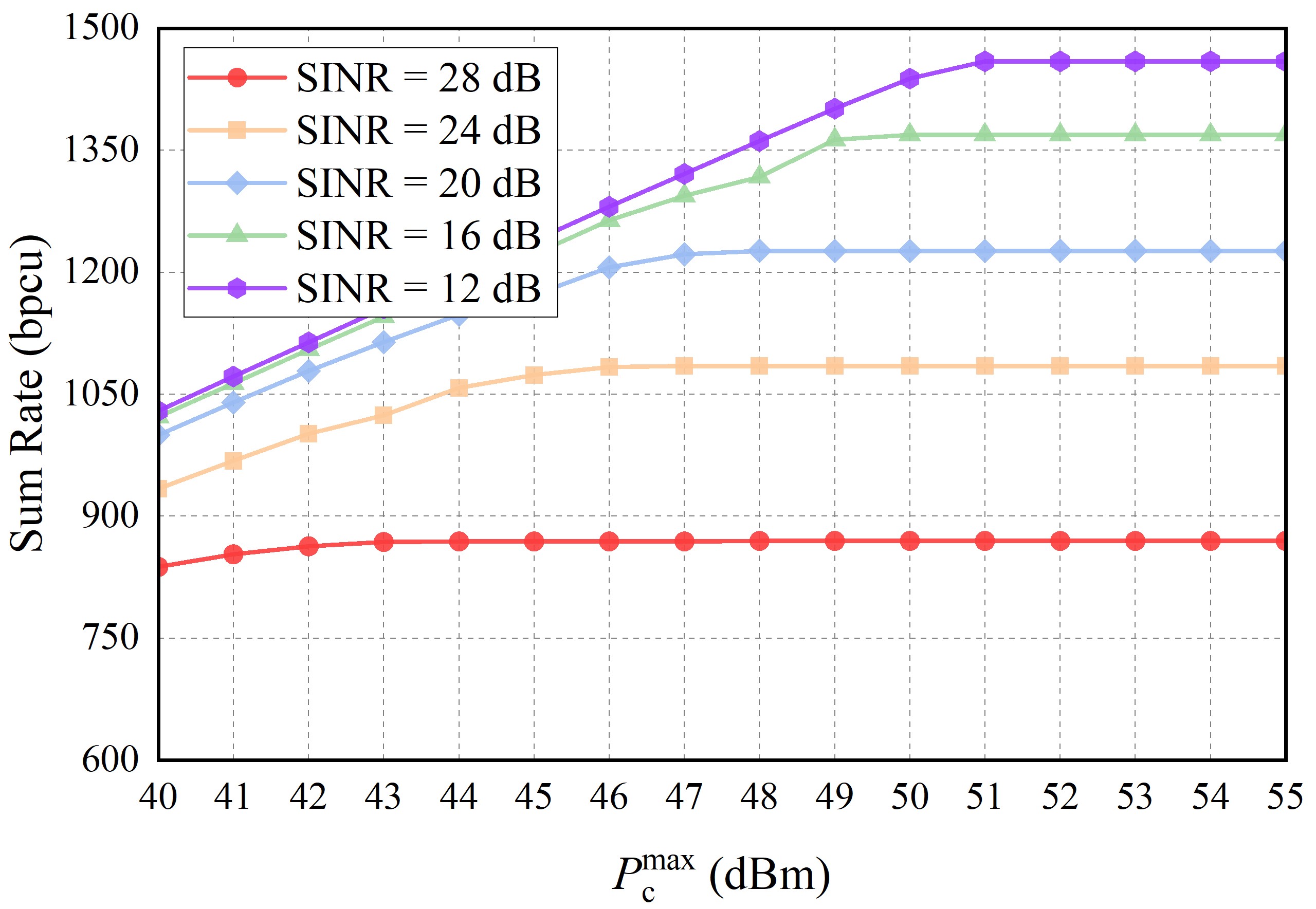}
	\caption{Sum rate versus the maximum power of the communication system with different radar SINR constraints.}
 \vspace{-3mm}
	\label{fig-3}
\end{figure}

\figurename~\ref{fig-2} shows the sum rate (in bits per channel use, bpcu) versus radar SINR when $P_c^{\max} = [40,42,44,46,48]$ dBm and $P_c = P_r =30$ dBm. According to the results shown in \figurename~\ref{fig-3}, the proposed algorithm is a nonlinear decreasing function of radar SINR. The reason for such a result is that the transmitted signal of the radar system will interfere with the communication system, and as the minimum SINR required by the radar system increases, the interference to the communication system will become more serious. The dotted line in the figure shows the communication rate in the absence of radar interference. It can be seen that when there is no radar interference, the sum rate is higher than the sum rate when the radar interference is present with same total communication power.
On the other hand, when the radar SINR becomes large, the total communication power has little effect on the total rate, and they tend to be the same.

\figurename~\ref{fig-3} shows the sum rate versus the total communication power when $\mathrm{SINR} = [12,16,20,24,28]$ dB and $P_r^{\max} = 45$ dBm. We observe that the sum rate increases as the total communication power increases under different SINR constraints. However, an interesting result is that although the total communication power is increasing, the sum rate converges to a constant beyond a certain power threshold. The larger the radar SINR is, the lower the threshold will be. This indicates that radar SINR constraints prevents the sum rate from increasing unboundedly as the total power increases. This result implies that the power of the communication system should be reasonably allocated under given radar SINR constraints to achieve power parsimony.

\begin{figure}[t]
	\centering
	\includegraphics[width=0.85\columnwidth]{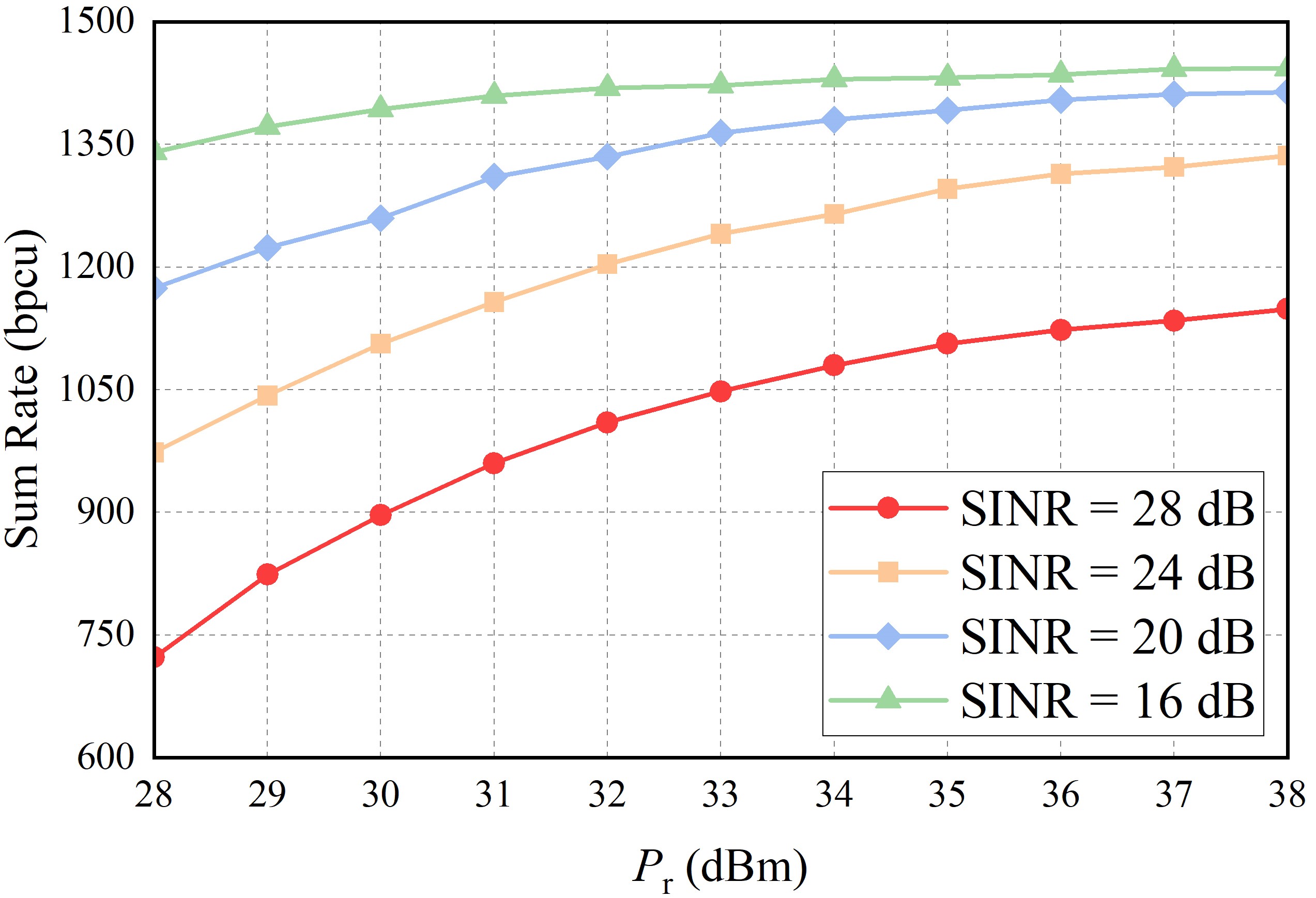} 
	\caption{Sum rate versus the maximum power of single radar subcarrier with different radar SINR constraints.}
    \vspace{-3mm}
	\label{fig-4}
\end{figure}

\figurename~\ref{fig-4} shows the sum rate versus the maximum per-subcarrier radar power $P_r$ when $P_c^{\max} = 50$ dBm and $P_c = 30$ dBm. \figurename~\ref{fig-4} indicates that the maximum power of single radar subcarrier is positively correlated with the achievable sum rate of communication system, which is the case under different SINR constraints. The conclusion that can be drawn is that the smaller the constraint on radar SINR, the weaker the impact of the change in maximum power of single radar subcarrier on the sum rate. 

\begin{figure}[t]
	\centering
	\includegraphics[width=0.85\columnwidth]{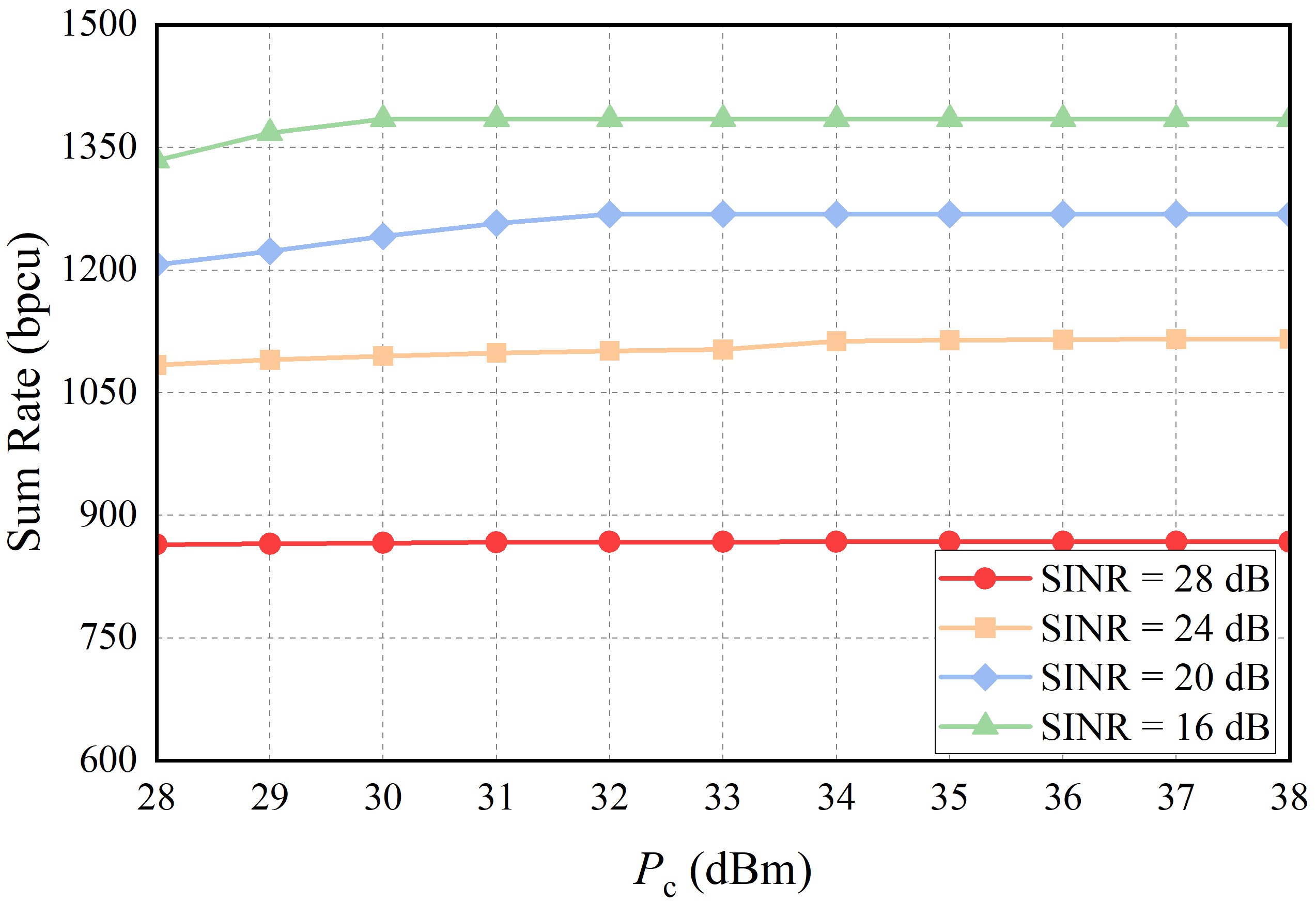} 
	\caption{Sum rate versus the maximum power of single communication subcarrier with different radar SINR constraints.}
 \vspace{-3mm}
	\label{fig-5}
\end{figure}

In contrast to \figurename~\ref{fig-4}, \figurename~\ref{fig-5} shows the impact of changing the maximum transmit power of a single communication subcarrier on the sum rate while keeping the remaining variables fixed. Overall, the change in the maximum transmission power of a single communication subcarrier has less impact on the sum rate compared to the impact of changing the maximum transmission power of a single radar subcarrier.
\section{Conclusions}\label{sec:Conclusions}
In this paper, we have investigated the power allocation problem in the spectrum coexistence of radar and communication systems, where we jointly allocate the communication transmission power and radar transmission power to maximize the sum rate of CUs under the constraint of radar sensing performance. Through proper reformulation, the problem containing binary variables is transformed into an equivalent optimization problem with only continuous-valued variables, and then the computationally tedious alternating optimization is replaced by an FP optimization in vector form. Simulation results exhibit the effectiveness of the algorithm and show the trade-off between communication rate and radar SINR. Especially, the interesting result that the sum rate does not increase with the total power beyond certain thresholds can be useful for the design of energy-efficient RCC systems.

\bibliographystyle{IEEEtran}
\bibliography{references}
\vspace{11pt}

\vfill
\end{document}